# Biological Tissue Imaging with a Position and Time Sensitive Pixelated Detector

Julia H. Jungmann[1], Donald F. Smith[1], Luke MacAleese[1,+], Ivo Klinkert[1], Jan Visser[2], Ron M.A. Heeren[1,*]

[1] *FOM Institute for Atomic and Molecular Physics (AMOLF), Science Park 104, 1098 XG Amsterdam, The Netherlands*

[2] *National Institute for Subatomic Physics (Nikhef), Science Park 105, 1098 XG Amsterdam, The Netherlands*

*+ current affiliation: Universite Lyon 1 and CNRS, UMR5579, LASIM, Villeurbanne, France.*

* Address reprint requests to Ron M.A. Heeren, FOM Institute for Atomic and Molecular Physics (AMOLF), Science Park 104, 1098 XG Amsterdam, The Netherlands, Tel: +31-20-7547 161, Fax: +31-20-7547 290, E-Mail: heeren@amolf.nl

**Abstract**

We demonstrate the capabilities of a highly parallel, active pixel detector for large-area, mass spectrometric imaging of biological tissue sections. A bare Timepix assembly (512 × 512 pixels) is combined with chevron microchannel plates on an ion microscope matrix-assisted laser desorption time-of-flight mass spectrometer (MALDI TOF-MS). The detector assembly registers position- and time-resolved images of multiple *m/z* species in every measurement frame. We prove the applicability of the detection system to bio-molecular mass spectrometry imaging on biologically relevant samples by mass-resolved images from Timepix measurements of a peptide – grid benchmark sample and mouse testis tissue slices. Mass-spectral and localization information of analytes at physiological concentrations are measured in MALDI-TOF-MS imaging experiments. We show a high spatial resolution (pixel size down to 740 × 740 $nm^2$ on the sample surface) and a spatial resolving power of 6 µm with a microscope mode laser field of view of 100 – 335 µm. Automated, large-area imaging is demonstrated and the Timepix' potential for fast, large-area image acquisition is highlighted.

**Introduction**

Mass spectrometry imaging (MSI) [1-4] measurements target the identification and localization of molecules from complex surfaces [5-11]. Over the past decade, MSI has developed into a widely used analytical tool for bio-molecular and bio-medical research. This is achieved as a result of technological developments and improvements in the areas of high spatial resolution, high mass accuracy and high throughput [12]. Matrix-assisted laser desorption/ionization (MALDI) time-of-flight (TOF) MSI is commonly used for bio-molecular and bio-medical studies because of the possibility for high spatial resolving powers, a wide accessible mass range at high acquisition speeds.

There are two major types of MALDI-TOF MSI; microprobe and microscope mode experiments. At present, commercial MALDI-MSI instruments for high throughput microprobe MSI studies are readily available. In MALDI microprobe MSI, a highly focused pulsed desorption/ionization laser beam probes the sample surface. A full mass spectrum is generated at every raster point. An image is reconstructed after the experiment with a pixel resolution equivalent to the beam size (provided the stepping size/accuracy of the sample/beam movement is not limiting). Highly focused ultra-violet/infra-red laser beams [6-8,11] have returned pixel sizes below 7 µm. In particular, the Spengler group has achieved a spatial resolution of less than 1 µm with a tightly focused ultra-violet laser beam [13]. Disadvantages of small surface probe areas include long measurement times and loss of sensitivity due to low laser fluence for desorption and ionization. MALDI microprobe MSI at a lateral resolution better than the laser spot size has been demonstrated by over-sampling [14].

Recent studies in microprobe mode MSI have focused on the improvement of sample throughput to obtain more practical large-area or spatial resolution conditions. McDonnell and co-workers use an automatic sample loading system for time-efficient, round-the-clock spectrometer operation [15]. Alternatively, high throughput can be achieved by an increase in the raster/laser frequency. This results in a decrease in the measurement time without sacrificing analytical information. The Caprioli group has reduced the measurement time by up to a factor of two by operating a MALDI-TOF MSI system in a continuous scanning mode at a comparatively high laser repetition rate of about 5 kHz [16]. Stoeckli and co-workers measured images of drug distributions in rat sections within less than 15 minutes with a 1 kHz laser repetition rate and a raster speed of about 18 mm/s [17].

MALDI microscope mode MSI is an alternative approach to microprobe mode imaging [5,18]. In the microscope mode, a large surface probe desorbs and ionizes surface molecules over a large sample area, typically 200-300 µm in diameter. The large field of view in microscope mode MSI enables fast, large area image acquisition at high spatial resolution [19]. An ion microscope uses ion optics to project the desorbed/ionized surface compounds onto a position-sensitive detector. Importantly, the initial ion distribution is magnified and the lateral spatial organization of the surface molecules is retained on the flight path to the detector. While the ionization spot size determines the obtainable spatial resolution in microprobe mode, the spatial resolution in microscope mode is decoupled from the ionization spot size. Rather, the spatial resolution is determined by the quality and the capabilities of the ion optics in combination with a position- (and time-) sensitive detector. Spatial resolving powers on the order of 4 µm have been achieved where a 10 µm charge-coupled device (CCD) camera pixel probed 1 µm on the sample surface [5,20].

A large desorption/ionization spot size ensures high throughput in microscope mode MSI. The microscope mode ionization beam size of about 200-300 µm is about one hundred times larger than a typical high spatial resolution microprobe mode desorption/ionization spot size. Hence, analyzing an area of 200 × 200 µm$^2$ corresponds to one measurement in microscope mode MSI. In microprobe mode, the analysis of the same area corresponds to 40,000 measurements, which would be significantly more time consuming at the same experimental repetition rate. The large ionization spot in microscope mode MSI can result in large ion loads on the detection system, which can lead to detector saturation. As a remedy, the ion generation is reduced (by, for instance, decreasing the laser power) and the integration time is increased to ensure adequate ion imaging. This effect can be counteracted with high desorption/ionization repetition rates and detector technology that can accommodate high ion loads. MALDI-TOF ion microscopes enable fast, automated, high resolution large area imaging provided that adequate, i.e. position- (and possibly time-) sensitive and fast detector systems are used to record high quality molecular images [19].

Many ion microscope implementations project the ions on a position-sensitive detector assembly, which is traditionally a micro-channel plate (MCP) followed by a phosphor screen and a CCD camera system [5] or a resistive/ delay-line anode [19,21]. The MCP, phosphor screen, CCD camera detection system cannot link the ion arrival time (i.e. the molecular weight in TOF MSI) to the spatial distribution. Typically, different molecular masses cannot all be measured in one acquisition but must be mass-selected for imaging. The information on the sample composition and associated spatial distribution is obtained by combining several mass-selected images from separate measurements. Delay-line detectors deliver space- and time-resolved information. Additionally, they have the advantage of precise ion arrival time measurements [22]. Alternatively, techniques are available for arrival time measurement via a decoupling circuit from the MCP stack [23]. Delay-line detectors still lack sufficient multi-hit capabilities for MALDI-MSI. Also, the reconstruction of the mass-resolved images is time-consuming, which renders ion optical tuning difficult due to the lack of direct image feedback.

Imaging detector technology has acquired an important role in biological, bio-molecular and bio-medical MSI. In-vacuum pixel detectors, which allow for position- [24,25] and time-resolved [25,26,20] photoelectron and ion imaging, have been increasingly used in the past few years. Brouard and co-workers have implemented an adapted velocity map imaging apparatus in combination with a fast framing CCD camera for microscope mode MSI [27]. In this alternative setup, a detector pixel probes 56 µm on the sample surface and a mass resolution of 5 Da was reported in the mass range from 300 – 500 *m/z.*

Recently, we demonstrated the successful implementation of MCP-Medipix2 [25] and MCP-Timepix [20] imaging detector assemblies for MSI at a spatial resolving power of few micrometers. In these experiments, a fully-integrated, solid state pixel detector of the Medipix/Timepix [28-30] detector family was combined with a chevron MCP and installed on an ion microscope imaging mass spectrometer. The major characteristics of this application-specific integrated circuit (ASIC) are 256 × 256 pixels of 55 × 55 μm$^2$ each per chip, no electronics noise and per pixel functionality. On the pixel level, the Timepix detector can be programmed to return both the position of impact of a particle (via the pixel address) and the time-of-flight of this particle (with respect to an external trigger, with a maximum time resolution of 10 ns). Thereby, the Timepix detector eleviates the need of mass-selection and repetitive measurements that were required with systems that could not measure the time-of-flight (i.e. an MCP-phosphor screen assembly). In a single experiment, mass-resolved images were acquired using this new setup [20]. These earlier experiments focused on the imaging capabilities and image quality of the Medipix2 pixelated detector for MSI [25]. Further, the mass spectral capabilities of the Timepix system were investigated, with respect to the detectable mass range and the sensitivity [20]. It was demonstrated that the pixelated detector unit facilitates the parallel detection of multiple molecular species and significantly reduces the measurement time for large area high resolution molecular imaging experiments.

The present work discusses the first results of an MCP-Timepix detector assembly for large-area, high resolution biological tissue imaging on a MALDI-TOF ion microscope mass spectrometer. In particular, the spatial resolution, the image quality and the analytical capabilities of an MCP-Timepix assembly in biological MSI are determined and discussed. The implementation of automated, large-area microscope mode MSI is demonstrated on mouse testis tissue sections. Large areas of tissue are rastered by the microscope beam and individual images (from each microscope mode position) are stitched together. The Timepix' potential for fast, large-area image acquisition is highlighted.

**Methods**

***The Medipix/ Timepix Detectors.*** The Medipix/ Timepix detector family is developed by the Medipix collaboration hosted by CERN [31,32]. The Medipix [28,29]/Timepix [30] read-out chips are designed in complementary metal oxide semiconductor (CMOS) technology. A single chip has a size of about 1.4 × 1.6 cm$^2$. The pixel matrix contains 256 × 256 pixels of 55 × 55 µm$^2$ each. Each application-specific integrated circuit (ASIC) chip is three-side buttable. Larger chip arrays of 2 × 2n chips (n=1,2, …) can be tiled with minimal dead space between the chips.

Typically, the Medipix2 chips [28,29] are combined with a semiconductor sensor layer such as silicon or cadmium telluride by bump-bonding every ASIC pixel to an equally-size pixel of the semiconductor sensor layer. For X-ray photon and electron detection, this semiconductor detection medium converts the particle impact energy into an electron-hole pair current, which is collected by one or multiple pixels. The number of charge carriers created by the particle's impact is proportional to the energy of the incident particle. Ions are typically not sufficiently energetic to penetrate into the sensor layer through the few hundred nanometer thick Aluminum bias contact. Ions can be detected indirectly by placement of a microchannel plate (MCP) in front of the detector [33,25,20]. Here, a so-called bare chip is combined with a MCP. The ASCIC's pixels register the electron clouds which an ion impact creates at the backside of the MCP. The use of a bare chip improves the response to electron showers through reduced in-sensor electron diffusion.

The Timepix chip is the successor chip of the Medipix2 chip. It is identical in dimensions and geometry to its predecessor. However, the two ASICs differ in their capabilities on the pixel level. The Medipix chip operates in single particle counting mode. Each pixel counts the number of events that impinge on the pixel. The Timepix ASIC's pixels can individually be selected to function in one of three modes: particle counting mode, time-of-flight (TOF) mode and time-over-threshold (TOT) mode. In the TOF mode, the arrival time of an event is measured with respect to an external trigger/shutter signal. In the TOT mode, the time is measured during which the charge resulting from the event exceeds the detection threshold level. In TOF and TOT modes, the maximum measurement duration is determined by the pixel counter depth, in combination with the measurement clock speed. The current maximum clock speed of 100 MHz results in a maximum measurement interval of 118 µs, which captures part of the ion mass spectrum (0 – 1650 Da in the low mass range, see Online Resources for further detail).

The Timepix assembly is read-out by a gigabit per second (1 Gbit/s) read-out system [34,35] developed within the Relaxd project (high Resolution Large Area X-ray Detector) [36]. This read-out is dedicated to 2 × 2 chip modules. It reads out the four chips in parallel and achieves frame rates of up to 120 frames/s. The user interacts with the chips via the dedicated acquisition software and graphical user interface "Pixelman" [37]. For details on the Medipix/ Timepix chips, available read-out interfaces and various applications the interested reader is referred to [32] and references therein. The interested reader is referred to our initial publication [20] and [26] for more detail on the technical aspects of the TPX detection system and it's implications for MSI.

*Imaging setup on the ion microscope.* The TOF ion microscope is a TRIple Focusing Time-of-flight (TRIFT II) mass spectrometer (Physical Electronics, Inc., Chanhassen, USA). The spatial resolving power of the mass microscope is determined mainly by a combination of the ion optical aberrations, the ion optical magnification factor and the capabilities of the imaging detector. The TRIFT ion microscope returns an ion optical magnification of about 40 - 100x. Therefore, a 55 × 55 µm$^2$ Timepix detector pixel corresponds to areas of 1.38 × 1.38 µm$^2$ or 550 × 550 nm$^2$ on the sample surface. Note that the latter resolution is smaller than the spatial resolving power that is accessible with the current setup due to ion optical aberrations [38]. All measurements are conducted in positive ion mode. The MALDI surface probe is the third harmonic (355 nm) of a neodymium-doped yttrium aluminum garnet (Nd:YAG) laser (Wedge, Model 355; Bright Solutions, , Cura Carpignano, Italy). The repetition rate is 10 Hz (pulse energy of 13 µJ, beam diameter 100 – 300 µm). This mass spectrometric ion microscope is described in detail elsewhere [5].

A 2 × 2 bare Timepix ASIC assembly is mounted 2mm behind a chevron MCP stack (F2225-21N290, Hamamatsu Photonics Deutschland GmbH, Herrsching am Ammersee, Germany; $\phi$ = 4 cm, 12 μm pores, 15 μm pitch). The reader is referred to [20] for details of the system. The MCPs are biased at a voltage of 1.85 kV (standard sample) and 2.25 kV (tissue sample), which corresponds to an MCP gain of $10^6$ and $5 \cdot 10^7$, respectively. The voltage between the MCP back and the chips was set to 0.6 kV. Typically the charge of one MCP electron shower is divided among multiple Timepix pixels (typically about 50 pixels). This effect and its relation to the dynamic range of the detection system were discussed previously [20].

***Timepix-generated spectra.*** In this work, the Timepix ASIC is operated exclusively in TOF mode. The master trigger of the TRIFT MS ion microscope triggers both the detection system and the MALDI laser so that one measurement frame of the Timepix detector corresponds to the image and spectral information from one laser shot (the trigger scheme was described previously [20]). The 512 × 512 (262 ,144) pixels act as individual parallel detectors, which each deliver the position (x- and y- pixel coordinate) and the TOF of an event per measurement frame (i.e. the Timepix pixels are single-stop time-to-digital converters). The data of every measurement frame is saved in a separate sparse data file that contains the pixel address and the corresponding TOF. Timepix acquired TOF spectra are built by filling the TOF values of several measurement frames into a histogram. This TOF spectrum is mass-calibrated using calibration coefficients calculated from a standard mixture sample. The maximum TOF interval of 118 μs, its relation to the accessible mass range and how wide mass range measurements can be set up by stitching together shorter spectral acquisitions have been discussed in reference [20].

***Timepix-generated large area images.*** Total ion images are generated by summation of multiple single-frame position measurements. Mass-selected images are generated by extracting the pixel positions corresponding to a particular mass peak (i.e. TOF) from the data set. Large-area "mosaic mode" images are generated by probing a macroscopic area on the sample surface with the diffuse ($\phi$=100 – 335 µm) laser beam as a surface probe. For example, an area of about 2 × 2 mm is analyzed in 32 × 32 "mosaic" tiles. Per mosaic tile, the images/TOF information corresponding to 150-300 laser shots are collected. Then, the stage is moved by 62.5 µm (2 mm/32) and the next acquisition is started via the TRIFT II control software (Wincadence, ersion 4.4.0.17, ULVAC-PHI, Inc., 2008). The MALDI laser used here has an elliptical shape, thus the oversampling between the acquisition of tile n and n+1 ensures that the full sample surface is probed. A home-built software package developed in Matlab (Matlab, Mathworks, Natick, USA; version 7.110.584, R2010b, 32-bits) stitches the mosaic images into a large area image by partially overlapping the images of the individual tiles. In the "overlap region" pixel intensities are corrected for double-counting by averaging. The program converts the data set into the format of the AMOLF Datacube Explorer [39], which is used for data analysis. Larger surface areas can be analyzed using larger mosaics or by stitching together a number of smaller area mosaics.

**Samples.** All samples were prepared on indium-tin-oxide (ITO) coated glass slides. A grid benchmark sample was prepared to assess the spatial resolving power of the system. A nickel hexagonal thin bar transmission electron microscopy (TEM) grid (700 mesh, G2760N, 3.05 mm diameter, 37 µm pitch, 8 µm bar width; Agar Scientific Limited, Stansted, UK) was placed on top of a 2 µl droplet of 1:1 (by volume) mixture of Bruker Peptide Calibration Standard II (Online resource 1, Table1; Bruker GmbH, Bremen, Germany) and 2,5-dihydroxybenzoic acid (DHB; 20 mg/ml in 50% methanol, 50% water, 0.02% TFA).

Mouse testicle (Harlan Laboratories, Boxmeer, The Netherlands; male balb/c mouse) were chosen as biologically relevant benchmark tissue samples. The mouse testis was cryo-sectioned to 20 µm thickness and thaw mounted on an ITO slide. The tissue slices were optically scanned prior to the MS experiments (Mirax Virtual Slide Scanner, Carl Zeiss AG, Oberkochen, Germany). To ensure small matrix crystals (Online Resources, Figure S1), the tissue was coated with a homogeneous layer of DHB matrix (10 mg/ml DHB in 50% methanol, 50% water, 0.02% TFA) with a home-built pneumatic spray device similar to that described by [40] (Figures 3-5). For large area MS imaging (Figure 6), DHB (20 mg/ ml DHB in 50% methanol, 50% water, 0.02% TFA) was applied with the Bruker ImagePrep, which allows fast, large area matrix coverage at matrix solution droplet sizes of ~ 50-80 µm (Online Resources, Figure S2).

**Results and Discussion**

*Imaging capabilities.* The peptide-TEM grid sample is used to test the large area imaging capabilities. In particular, the image quality of the stitched mosaic microscope mode image and the spatial resolution is assessed. In this experiment, an automated microscope mode acquisition of 32 × 32 tiles is used. The ion microscope probes an area of 2.24 × 2.24 mm$^2$ with a raster of 70.1 µm between tiles at a magnification factor of about 62. At this magnification factor, each Timepix pixel probes 880 nm on the sample surface. One hundred individual Timepix acquisition frames corresponding to 100 laser shots are acquired per position and integrated offline. The peptide standard spectrum of this experiment contains the expected peptide peaks in the mass range 700 < *m/z* < 3500 (Figure S3, Online Resources). The spectral quality is in accordance with the time resolution of the Timepix detections system and has been investigated and discussed in detail elsewhere [20]. Figure 1 displays the stitched "mosaic" images of a microscope mode MSI experiment on the peptide standard – TEM grid sample at four different pixel sizes. At the full spatial resolution, i.e. when one detector pixel probes 880 × 880 nm$^2$ on the sample surface, the image is composed of 3147 × 3147 pixels (Fig. 1a). This number of pixels is about 7 times higher than the number of pixels on a standard modern computer screen and corresponds to about 28,800 dots per inch. Hence, to fully appreciate the amount of detail revealed by the high spatial resolution, part of the mosaic image can be zoomed into (see Fig. 2). The inhomogeneity of the intensity across the stitched total ion image originates from inhomogeneities in the amount of analytes located in the different areas underneath theTEM grid. The high resolution, large-area microscope mode image obtained at Timepix resolution (Fig. 1a) is contrasted with images of the sample at lower spatial resolutions. Figure 1 also displays the stitched mosaic image at resolutions where one pixel probes 18 × 18 μm$^2$ (Fig. 1b), 60 × 60 μm$^2$ (Fig. 1c) and 119 × 119 μm$^2$ (Fig. 1d) on the sample surface. These images are obtained by spatially binning the Timepix acquired data. In particular, one pixel probes 18 × 18 μm$^2$ if the counts of 20 × 20 Timepix pixels are binned into a single pixel. Similarly, one pixel probes 60 × 60 μm$^2$ or 119 × 119 μm$^2$ on the sample surface if the counts of 68 × 68 and 135 × 135 Timepix pixels are binned into a single pixel. The comparison between the different spatial resolutions demonstrates the increase of spatial detail that is revealed with the increase of the spatial resolution. The interested reader is referred to the Online Resources for a comparison of the Timepix microscope mode image (Figure 1) to a microprobe mode image of a peptide-TEM grid sample (Figure S4) measured with a SYNAPT G1 HDMS (Waters Corporation, Milford, USA).

The peptide mixture underneath the TEM grid is clearly observed in the image at full Timepix resolution (880 nm) and with somewhat less detail at the 18 × 18 μm$^2$ pixel size. At the pixel sizes of 60 × 60 μm$^2$ and 119 × 119 μm$^2$, the image spatial resolution ranges in regime of standard, commercial MALDI microprobe instruments. At these pixel sizes, the spatial resolution is insufficient to reveal the structure of the TEM grid. The signal intensities displayed in Figure 1 represent absolute measurement intensities and are not scaled. As expected, the image contrast is better at larger pixel sizes (maximum intensity = 926,268) than at full resolution (maximum intensity = 258) since the same number of ion counts on the detection system is displayed in less spatial bins.

Note, that the intensity values represent relative intensities. No conclusions should be drawn from the absolute intensity values delivered by the Timepix detection system. The Timepix pixels are single-stop time-to-digital converters, i.e. each pixel can accommodate one count per measurement frame and is blind for other ions to arrive later in time-of-flight during the same measurement frame. Operation in a moderate count rate regime minimizes this effect. Figure 2 is used to assess the image spatial resolution at full Timepix resolution, i.e. 1 detector pixel probes 880 × 880 nm$^2$ on the sample surface. Figure 2A displays a line scan through part of the high resolution image. Figure 2B shows a zoom on part of the full resolution Timepix image, where the position of the line scan is indicated. The line scan spans a distance corresponding to 74 µm on the TEM grid sample such that two grid cells (pitch 37 µm) are included. The spatial resolving power, i.e. sharpness of a feature edge, has previously been defined as the distance between 20% to 80% signal intensity of a feature within an image [1]. With this criterion, a spatial resolving power of ~ 6 µm is achieved with the ion microscope in combination with the Timepix detection system. A spatial resolving power on this order of magnitude is expected (i.e. not 880 nm) due to the need for multiple pixels to resolve the feature. Specifically, the MCP electron showers create signal clusters of about 6 - 10 pixels in diameter on the Timepix detector. Such a cluster corresponds to (cluster diameter · pixel size)/(magnification factor) µm on the sample surface, i.e. (6 · 55 µm)/62 = 5.3 µm to (10 · 55 µm)/62 = 8.9 µm. A decreased cluster size would result in a higher spatial resolving power. In addition, large DHB matrix crystals due to the dried droplet sample application can decrease the image resolution. To fully exploit the benefit of the high spatial resolution, significant sample preparation effort is required to ensure that the spatial resolution is not hampered by the sample treatment. This would occur if the matrix crystal size was significantly larger than the achievable pixel size is [41]. The subtle vertical and horizontal lines in Fig. 2b represent artifacts from the stitching of the mosaic tiles (one tile is emphasized by the broken white lines) and can be removed using smoothing algorithms that have not been applied in these raw images. Note that the borders of these lines define the area of a single measurement tile. Also note that the peptide mixture deposited underneath the metal TEM grid unlike the metal grid itself is readily ionized in this MALDI-MSI experiment, which leads to broadening of this image feature.

***Spectral capabilities on tissue.*** The spectral capabilities of the detection system on biological tissue samples are investigated with the mouse testicle sample. Figure 3 shows a mass spectrum from 93,000 integrated Timepix acquisition frames (i.e. 93,000 laser shots) at a spectral bin size of 1 Da (the interested reader is referred to [20] for a detailed evaluation of the spectral capabilities of the MCP/Timepix detection system). The spectrum is obtained by integrating eight selected-area images that are similar to the two images displayed in Fig. 4. This number of acquisition frames corresponds to probing about 7% of the testis surface area. The total ion spectrum was baseline subtracted with a linear regression method (window size 5 points, step size 3 points). Ions between 700 *m/z* and 900 *m/z* can be attributed to commonly observed lipid species, which demonstrates that spectral analysis at physiological analyte concentrations on tissue is feasible with this Timepix-based detection system. The inset of Fig. 3 displays the spectrum at 1500 < *m/z* < 7130. In this *m/z* region, distinct peaks are observed which shows that species beyond the lipid mass range are accessible with the Timepix-based detection system.

*Mass-selected imaging on a biological sample.* The imaging capabilities of the detection system are tested on the mouse testis sample. Figure 4 displays an optical scan of the tissue slice that is used for the MS imaging experiment. The sample was coated with DHB by a pneumatic sprayer to ensure small matrix crystals of ~10 μm in length. Two large area acquisitions are carried out at an ion optical magnification factor of 75. The experiment shown in Fig. 4b,c probes an area of 0.5 × 0.5 mm$^2$ with 8 × 8 microscope mode mosaic tiles. The tile-to-tile raster shift is 62.5 μm and 150 Timepix images (i.e. laser shots) are acquired per tile. The second experiment (Fig. 4d,e) images an area of 1 × 1 mm$^2$ with 16 × 16 microscope mode mosaic tiles. The tile-to-tile raster is 56.3 μm and 150 Timepix acquisition frames (i.e. laser shots) are collected. Fig. 4 displays the total ion image of these experiments (Fig.4b,d) and corresponding mass-selected ion images (Fig. 4c,e).

The images are displayed at the full Timepix resolution and are scaled to 25% of the maximum pixel intensity for increased image contrast. The mass-selected images clearly reveal the distribution of bio-molecular analytes at physiological concentrations from the tissue surface. The detection of these analytes at the current acceleration voltage of the ion microscope of 5 keV demonstrates the sensitivity of the detection system.

The total ion image of the 16 × 16 mosaic shows image artifacts due to detector saturation. The count rates in this MALDI experiment are high, which results in detector saturation effects that are observed in some tiles as the vertical flares in Fig. 4d. This detector saturation effect can be eliminated from the data set in two ways. First, the mass-selected images do not contain the saturation effects. The pixels that are in saturation do not return valid TOF/mass spectral information and can be excluded by this criterion. Secondly, the Timepix acquisition frames that contain saturated images contain significantly more pixel counts than the average MALDI frame, in which about 40-70% of the Timepix pixels are triggered. Hence, the corresponding data files are larger in size and can be excluded from further analysis by this criterion. Note that the detector saturation effect prevents the detection of ion events in this area of the detection system and hence distorts the ion optical image and relative image intensity.

Accurate peak assignment remains challenging with the Timepix detection systems in combination with the TOF mass spectrometer. The Timepix clock cycles of 10 ns result in wide mass bins and wide peaks in the spectrum. Thus, the current mass resolution does not enable analyte identification, as outlined in [20]. Successor chips with a higher time resolution will improve the capabilities for mass assignment significantly. Figure 4c shows mass-selected images with *m/z* = 491-495 (blue), *m/z* = 595-599 (green), *m/z* = 621-624 (red) for the 8 × 8 mosaic.  Figure 4e displays mass-selected images with *m/z* = 504-507 (green) and *m/z* = 621-624 (red) for the 16 × 16 mosaic. The analyte at *m/z* = 621-624 might be heme, since the optical image in Fig. 4a displays blood in the area where the analyte is localized in the mass-selected image. The walk "discrepancy" in the mass relates to the above mentioned limitations. The *m/z* = 595-599 analytes might be a di-acyl-glycerol.

***Advantages of Microscope Mode Imaging for High Spatial Resolution.*** The capabilities of high spatial resolution imaging are investigated in closer detail in the 8 × 8 microscope mode mosaic on the mouse testis sample (Fig. 4b,c). Figure 5 shows the total ion image and mass-selected ion images of this mosaic at different pixel sizes. At the ion optical magnification factor of about 75 and at the full Timepix resolution, one Timepix pixel probes 740 × 740 nm$^2$ on the sample surface (Fig. 5a, scaled to 25% of the image intensity for increased image contrast). The Timepix acquired data is spatially binned by combining the counts of 20 × 20 (pixel size: 15 × 15 μm$^2$, Fig. 5b), 68 × 68 (pixel size: 50 × 50 μm$^2$, Fig. 5c), and 135 × 135 (100 × 100 μm$^2$, Fig. 5d) pixels, respectively. The comparison between the different spatial resolution images demonstrates that accurate analyte localization is only ensured at the fine pixel pitches. In particular, the two distinct areas of the analytes at *m/z* = 595-599 (displayed in green) is not resolved by the pixel sizes of 50 × 50 μm$^2$ and 100 × 100 μm$^2$, respectively. The pixel size of 15 × 15 μm$^2$ and the pixel size of 740 × 740 nm$^2$ in even greater detail reveals the spatial structure of this analyte.

***Large-area imaging and throughput capabilities.*** Large-area tissue imaging typically targets areas of several square centimeters. Fast, automated MSI technology is required to ensure high throughput experiments. The Timepix detection system in combination with a high frame rate read-out system is a potential candidate for high throughput MSI experiments. In the proof-of-principle setup presented here, however, the full potential of the Timepix system cannot be exploited as it could be in a dedicated measurement system, due to data read-out limitations.

Figure 6 displays mass-selected images from a selected-area imaging experiment on the mouse testis sample. The field of view is 1.8 × 1.8 mm$^2$, which corresponds to about 11% of the area of the mouse testis section. This area is imaged in 32 × 32 microscope mode mosaic tiles at a raster of 56.3 μm. The ion optical magnification factor is 75 such that the pixel resolution corresponds to 740 × 740 nm$^2$ on the sample surface. Per position, 150 frames corresponding to 150 laser shots are acquired. This typical imaging experiment is based on a total of 32 × 32 × 150 = 153,600 laser shots. At a repetition rate of 10 Hz, this corresponds to a measurement time of 4 hours and 20 minutes. If the experimental repetition rate was 1 kHz, the measurement would only take 2.5 minutes. This means that the entire mouse testicle section (5.4 × 5.4 mm$^2$) could be measured within a microscope mode MSI experiment of about 23 minutes. For comparison, a microprobe experiment under the same experimental conditions takes significantly longer due to the small ionization probe size. An ionization probe of about 3 µm results in $(1.8 \text{ mm}/3 \text{ µm})^2$ = 360,000 measurement positions. Hence, the experiment requires about 360,000 × 150 = 54,000,000 laser shots and takes about 350 times longer than a microscope mode experiment.

The main limiting factor to the throughput of the system is the repetition rate which is determined by the repetition rate of the read-out system in combination with the read-out software package. Since the chip itself can, in principle, be readout within 300 µs (parallel readout through the 32-bit CMOS lines [30]), repetition rates of 1 kHz are feasible. Within the Medipix collaboration, several read-out interfaces are under development to specifically target high frame rate applications. For the Medipix2 chip, there are the "Parallel Readout Image Acquisition for Medipix" (PRIAM) and the "Dear-Mama Acquisition System" (DEMAS) systems for reading out frames at kilohertz [42] and 500Hz [43] frame rates, respectively. Recently, the Berkeley Quad Timepix Parallel Readout [44] was presented. It is a highly parallel read-out system dedicated to the read-out of 2 × 2 Timepix ASICs at a frame rate of 1 kHz. These read-out systems would enable high throughput MSI measurements for the Timepix system described herein. However, the applicability of these kHz read-out systems for MALDI-TOF count rates needs to be investigated.

**Conclusions and Outlook**

Mass spectrometry imaging (MSI) on biologically relevant samples was shown using a chevron microchannel stack in combination with a bare Timepix application-specific integrated circuit on an ion microscope mass spectrometer. Unlike commercial MSI detection systems based on a MCP, phosphor screen and charge-coupled device camera, this novel detection system provides two-dimensional imaging information, i.e. localization and arrival time measurements, i.e. the *m/z* of the ions, from a single detection system within every single measurement frame. Hence, the Timepix facilitates faster MSI measurements by removing the need for repetitive measurements on different *m/z* species.

The spatial resolution of this detection system on the ion microscope results in pixel sizes of 740 – 880 nm and a spatial resolving power of about 6 µm. The imaging capabilities on biologically relevant samples indicate the system's potential for bio-molecular and bio-medical MSI, since significantly more structure can be resolved with the Timepix system in microscope mode than at commercially available pixel sizes. Mass-selected images can be generated and the mass spectra show high signal-to-noise ratios. The mass resolution is in accordance with the expectations for a 100 MHz measurement clock.

The throughput of the current Timepix detector in large-area imaging is limited by the low repetition rate of the read-out system. Read-out systems that perform two orders of magnitude faster repetition rates are feasible and need to be tested at matrix-assisted laser desorption/ionization time-of-flight mass spectrometry (MALDI-TOF-MS) count rates. A Timepix successor chip (as sketched in [20]) in combination with a kHz read-out system and an ion microscope will result in a state-of-the-art microscope mode MALDI-TOF-MSI instrument. Such as system will possess a high dynamic range, a low signal-to-noise ratio, a high mass range, an extremely high spatial resolution and will be capable of the simultaneous measurement of both localization and arrival time of various m/z ions during one single measurement.

The presented work demonstrates the analytical capabilities of this novel detection system for mass spectrometry imaging on biologically relevant samples. Conveniently, the high degree of flexibility in the design of the MCP/Timepix detector assembly enables the implementation of this detection system on any time-of-flight mass spectrometer. The combination of the system's high spatial resolution capabilities with a high mass resolution, tandem MS or ion mobility mass spectrometer in the future will expand the high end toolbox for MS imaging.

**Acknowledgements**

Part of this research is supported by the Dutch Technology Foundation STW, which is the applied science division of NWO, and the Technology Programme of the Ministry of Economic Affairs, Project OTP 11956. This work is part of the research program of the "Stichting voor Fundamenteel Onderzoek der Materie (FOM)", which is financially supported by the "Nederlandse organisatie voor Wetenschappelijk Onderzoek (NWO)". The authors acknowledge Ronald Buijs, Frans Giskes, Marc Duursma, Bas van der Heijden, Marten Bosma and Joop Rövekamp for their contribution to the experiments.

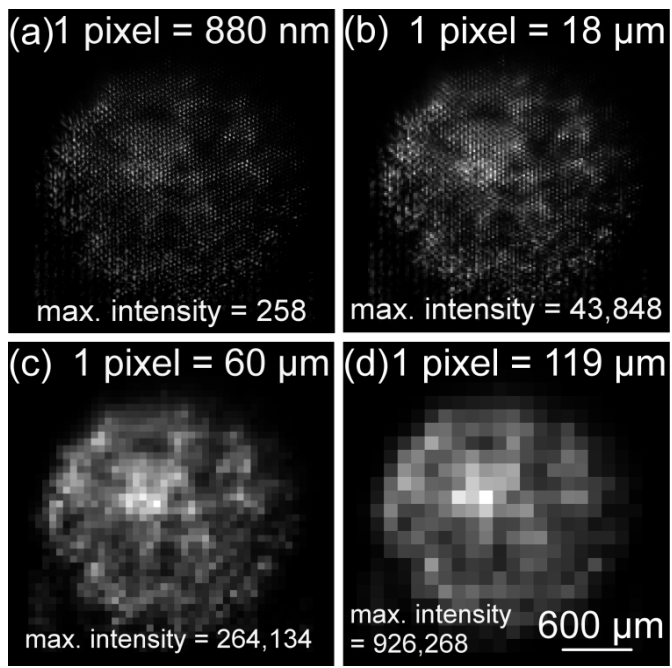

**Figure 1:** *Total ion image of the peptide standard – TEM grid sample. The image is composed by stitching together 32 × 32 microscope mode images. (a) full Timepix resolution, one detector pixel probes 880 × 880 nm$^2$ on the sample surface. (b,c,d) Images at different image resolutions obtained from the same experiment by spatially binning the Timepix data.*

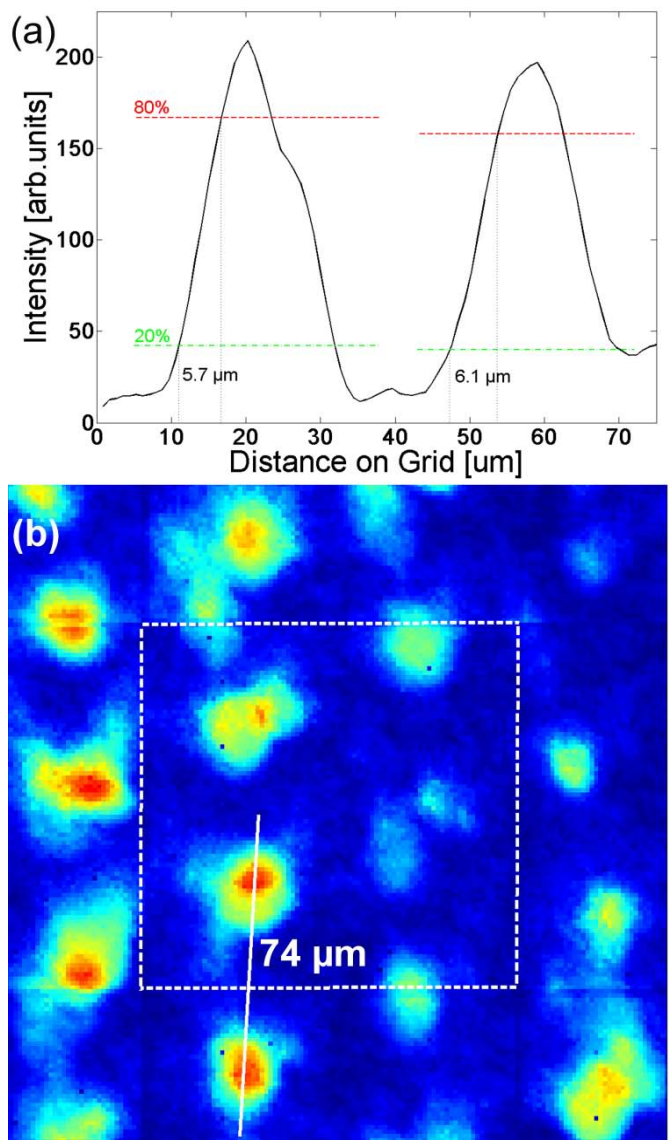

**Figure 2:** *(a) Line scan through the image of the peptide mixture – TEM grid sample to assess the spatial resolving power of the imaging setup. The red and green lines indicate 80% and 20% of the maximum peak intensity, respectively. The spatial resolving power is defined as the distance between these two points. (b) Zoom on part of the high resolution image of the peptide standard – TEM grid sample. One pixel in the image corresponds to 880 × 880 nm$^2$ on the sample surface. The line used to generate Fig. 2a is shown in white.*

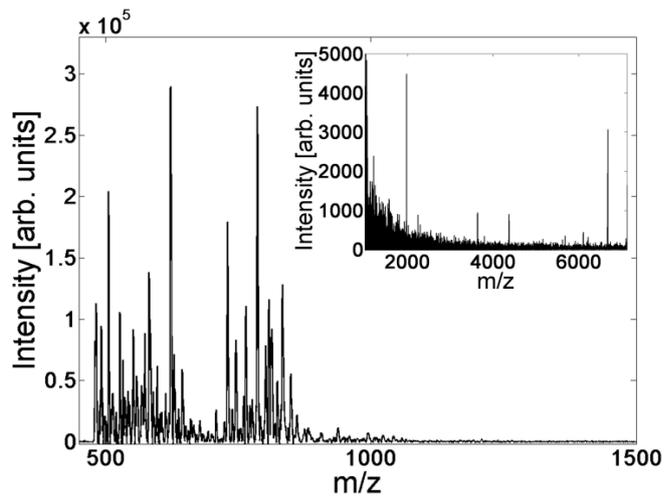

**Figure 3:** *Summed, baseline subtracted mass spectrum from 93,000 acquisition frames from mouse testis measured with the Timepix. Many ions are observed in the lipid mass range (700-900 m/z) and the zoom inset shows higher mass ions.*

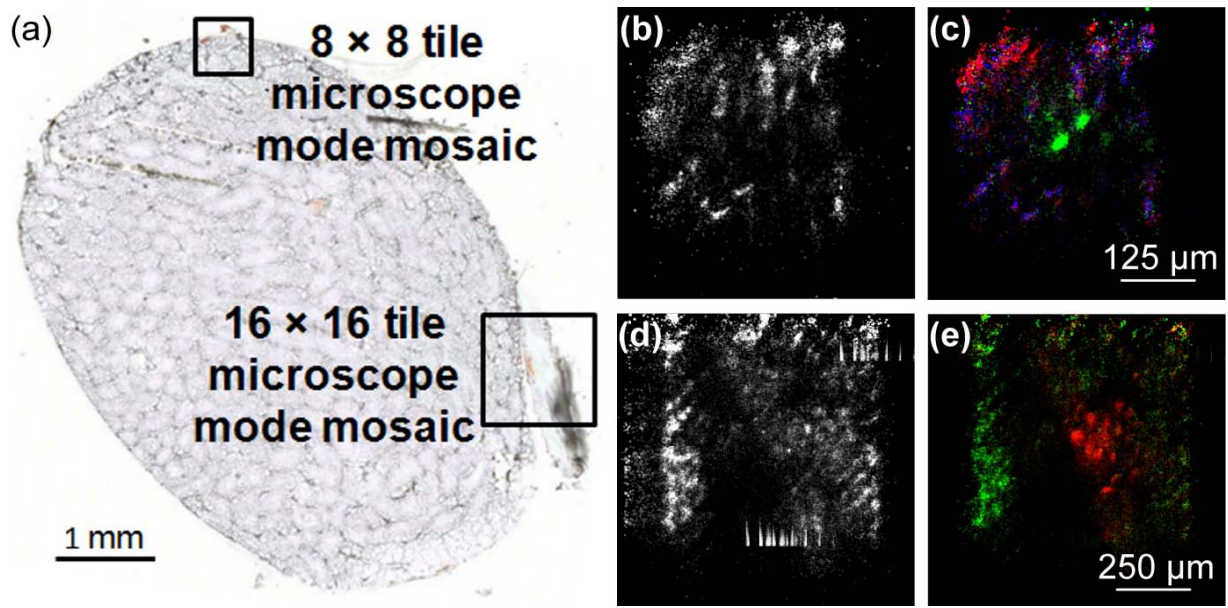

**Figure 4:** (a) *Scan of the mouse testis tissue sample. The indicated black squares indicate two areas which have been imaged with the Timepix detection system on the MS ion microscope. (b) Total ion image from the 8 x 8 (0.5 x 0.5 mm$^2$) mosaic, (c) mass-selected images with m/z = 491-495 (blue), m/z = 595-599 (green), m/z = 621-624 (red) for the 8 × 8 mosaic, (d) total ion image from the 8 × 8 (1 x 1 mm$^2$) mosaic and (e) mass-selected images with m/z = 504-507 (green) and m/z = 621-624 (red) for the 16 × 16 mosaic.*

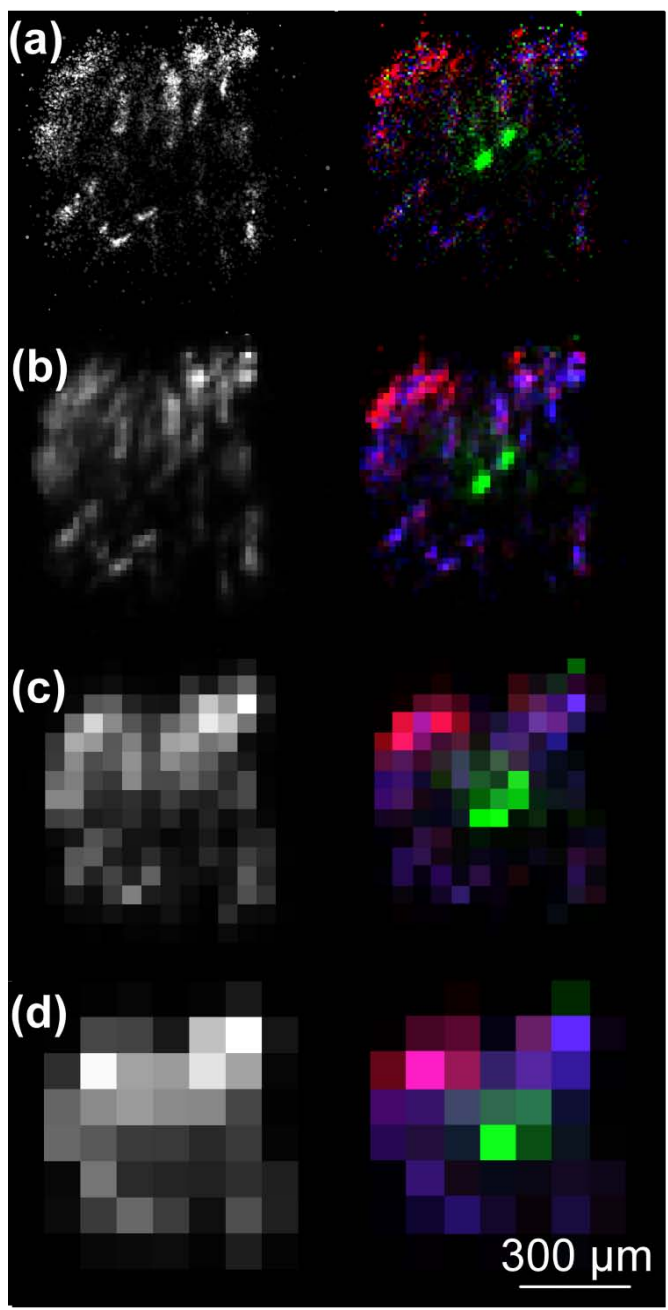

**Figure 5:** *Mouse testis selected-area images at different pixel sizes: 1 pixel probes (a) 740 nm, (b) 15 µm, (c) 50 µm and (d) 100 µm on the sample surface. (left) total ion images. (right) mass-selected images with m/z = 491-495 (blue), m/z = 595-599 (green), m/z = 624-624 (red).*

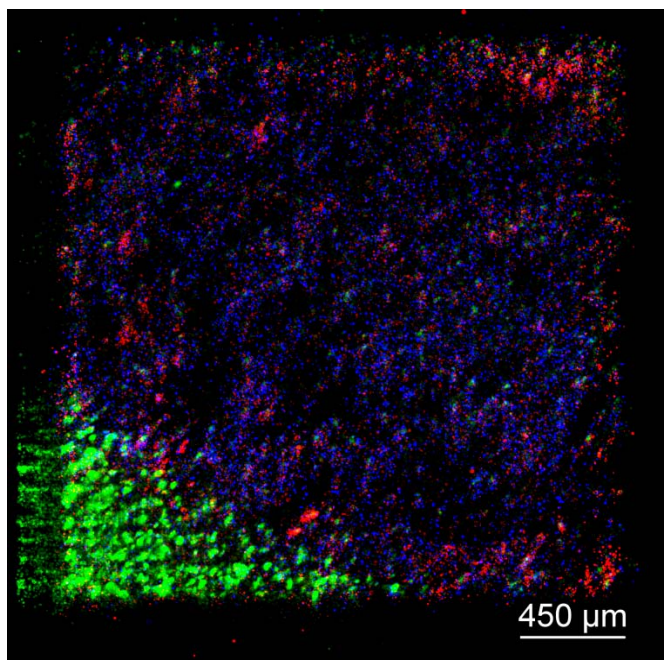

**Figure 6:** *Mass-selected mouse testis images at the full Timepix resolution. The field of view is 1.8 × 1.8 mm$^2$ (32 × 32 microscope mode mosaic tiles, 2868 × 2868 pixels) and one pixel probes 740 × 740 nm$^2$ on the sample surface. The selected analyte mass range of m/z = 320-322 (green) appears to be matrix-related since it is observed off-tissue. M/z = 618-621 (red) might correspond to heme. The analyte at m/z =726-729 (blue) is most likely a lipid. The mass range was blanked below m/z 300. (Note that the color scheme used here is not the same as in Figure 4 and 5.)*